# Superconductivity in Novel BiS$_2$-Based Layered Superconductor LaO$_{1-x}$F$_x$BiS$_2$

Yoshikazu Mizuguchi[1,2,*], Satoshi Demura[2], Keita Deguchi[2], Yoshihiko Takano[2], Hiroshi Fujihisa[3], Yoshito Gotoh[3], Hiroki Izawa[1], Osuke Miura[1]

[1]*Department of Electrical and Electronic Engineering, Tokyo Metropolitan University, 1-1, Minami-osawa, Hachioji, 192-0397, Japan*

[2]*National Institute for Materials Science, 1-2-1, Sengen, Tsukuba, 305-0047, Japan*

[3]*National Institute of Advanced Industrial Science and Technology (AIST), Tsukuba Central 5, 1-1-1, Higashi, Tsukuba 305-8565, Japan.*

Layered superconductors have provided some interesting fields in condensed matter physics owing to the low dimensionality of their electronic states. For example, the high-$T_c$ (high transition temperature) cuprates and the Fe-based superconductors possess a layered crystal structure composed of a stacking of spacer (blocking) layers and conduction (superconducting) layers, CuO$_2$ planes or Fe-Anion layers. The spacer layers provide carriers to the conduction layers and induce exotic superconductivity. Recently, we have reported superconductivity in the novel BiS$_2$-based layered compound Bi$_4$O$_4$S$_3$. It was found that superconductivity of Bi$_4$O$_4$S$_3$ originates from the BiS$_2$ layers. The crystal structure is composed of a stacking of BiS$_2$ superconducting layers and the spacer layers, which resembles those of high-Tc cuprate and the Fe-based superconductors. Here we report a discovery of a new type of BiS$_2$-based layered superconductor LaO$_{1-x}$F$_x$BiS$_2$, with a $T_c$ as high as 10.6 K.

## 1. Introduction

Layered crystal structures are an exotic stage to explore superconductors with a high transition temperature ($T_c$) and to discuss the mechanisms of exotic superconductivity, as have the high-$T_c$ cuprates[1-4] and the Fe-based superconductors.[5-13] The discovery of basic superconducting layers, such as the CuO$_2$ plane and Fe$_2$An$_2$ (An = P, As, S, Se, Te) layers, have opened new fields in physics and chemistry of low-dimensional superconductors, because many analogous superconductors can be designed by changing the structure of the spacer layers. Recently, we reported superconductivity in the novel BiS$_2$-based superconductor $Bi_4O_4S_3$.[14] The crystal structure analysis indicated that $Bi_4O_4S_3$ had a layered crystal structure with space group of $I4/mmm$. The structure is composed of a stacking of rock-salt-type $BiS_2$ layers and $Bi_4O_4(SO_4)_{1-x}$ layers (blocks), where $x$ indicates the defects of $SO_4^{2-}$ ions at the interlayer sites. Thus, the parent phase ($x = 0$) is $Bi_6O_8S_5$, and $Bi_4O_4S_3$ is expected to have about 50 % defects of the $SO_4^{2-}$ site ($x = 0.5$). The band calculations indicated that $Bi_4O_4S_3$ ($x = 0.5$) was metallic while the parent phase of $Bi_6O_8S_5$ ($x = 0$) was found to be a band insulator with $Bi^{3+}$. In $Bi_4O_4S_3$, the Fermi level lies within the bands which mainly originate from the Bi 6p orbitals. In particular, the Fermi level is just on the peak position of the partial density of states of the Bi 6p orbital within the $BiS_2$ layer. With respect to the fact that the $BiS_2$ layers mainly contribute to the superconductivity of $Bi_4O_4S_3$, we expected that the $BiS_2$ layer would be a basic structure for a new class of an exotic superconducting family. Therefore we investigated the electron doping effects in an analogous $BiS_2$-based layered compound $LaOBiS_2$.[15] The parent phase of $LaOBiS_2$ is a band insulator, and possesses a layered structure composed of $BiS_2$ layers and $La_2O_2$ layers like $Bi_6O_8S_5$. Here we report the evolution of superconductivity in the $BiS_2$-based layered compound $LaO_{1-x}F_xBiS_2$ by electron doping via F substitution at the O site. The appearance of superconductivity upon electron doping is corresponding to the case of $Bi_4O_4S_3$.

*E-mail: mizugu@tmu.ac.jp

## 2. Experimental details

Polycrystalline samples of $LaO_{1-x}F_xBiS_2$ were prepared using three different methods, AP1, AP2 and HP, where AP and HP stand for the second annealing at "ambient pressure" and "high pressure", respectively. All the chemicals used in this study were purchased from Kojundo Chemical Lab. For the method AP1, we prepared the polycrystalline samples using powders of $La_2O_3$ (99.9%), $LaF_3$ (99.9%), $La_2S_3$ (99.9%), $Bi_2S_3$ and Bi (99.99%) grains. The $Bi_2S_3$ powders were prepared by reacting Bi and S (99.9%) grains at 500 ºC in an evacuated quartz tube. The starting materials with a nominal composition of $LaO_{1-x}F_xBiS_2$ were well-mixed, pressed into pellets, sealed into an evacuated quartz tube, and heated at 800 ºC for 10 h. The product was ground, mixed for homogenization, pressed into pellets and annealed again in an evacuated quartz tube at 800 ºC for 10 h. For the method AP2, we prepared the polycrystalline samples using powders of $Bi_2O_3$ (99.9%), $BiF_3$ (99.9%), $La_2S_3$, $Bi_2S_3$ and Bi grains. The sample preparation process is almost the same as AP1, except for the heating temperature. Using this method, we could obtain single-phase samples with heating at 700 ºC. For the method HP, we prepared precursor powders of $LaO_{1-x}F_xBiS_2$ by first annealing with the process in AP2, and performed a second annealing at 600 ºC under a high pressure of 2 GPa using a cubic-anvil high pressure synthesis instrument with 180 ton press.

X-ray diffraction (XRD) patterns were collected by a RIGAKU x-ray diffractometer with Cu-K$\alpha$ radiation using the $2\theta$-$\theta$ method. Lattice parameters were calculated using the peak positions by least-square calculations. Rietveld refinements were performed using RIETAN2000 program.[16] In the refinement, the small impurity peaks near 27.5º and 43.5º were excluded. The schematic image of the crystal structure was depicted using VESTA.[17]

Temperature dependence of magnetic susceptibility from 15 to 2 K after both zero-field cooling (ZFC) and field cooling (FC) was measured using a superconducting quantum interference device (SQUID) magnetometer with Magnetic Properties Measurement System (Quantum Design). Temperature dependence of resistivity from 300 to 2 K was measured using the four terminals method with Physical Properties Measurement System (Quantum Design). The $T_c^{onset}$ was defined to be a temperature where the resistivity begins to decrease.

## 3. Results

*3.1 Structural analysis of $LaO_{1-x}F_xBiS_2$.*

LaOBiS$_2$ has a layered crystal structure with a space group of *P*4/*nmm*. Figure 1(a) displays a schematic image of the crystal structure of LaOBiS$_2$. The structure is composed of a stacking of La$_2$O$_2$ layers and Bi$_2$S$_4$ layers (two BiS$_2$ layers in the unit cell), which is analogous to the Bi$_4$O$_4$S$_3$ superconductor (see Fig. 1(b)). To dope electrons into the BiS$_2$ conduction layers, we substituted O$^{2-}$ by F$^-$ with a range of $x$ = 0 ~ 0.7 in LaO$_{1-x}$F$_x$BiS$_2$. Figure 2(a) shows the x-ray diffraction (XRD) patterns for $x$ = 0 and 0.5 synthesized using three methods of AP1, AP2 and HP (see Methods). Almost all of the observed peaks are well indexed to the space group of *P4/nmm* (Fig. 2(b)), except for a few peaks relating to impurity phases of BiF$_3$ or LaF$_3$ near 27.5 º and 43.5º. For $x \geq$ 0.4, impurity peaks appear, indicating the existence of a solubility limit of O/F near $x$ = 0.4. It is found that the XRD profiles for the LaO$_{0.5}$F$_{0.5}$BiS$_2$ samples prepared using two different methods of AP1 and AP2. For the LaO$_{0.5}$F$_{0.5}$BiS$_2$ sample prepared by HP, the broadening of the peaks is observed, which implies that LaOBiS$_2$ system is not stable at high pressures. To confirm the crystal structure of the F-substituted system, we performed Rietveld refinement. Figure 2b shows the result of Rietveld refinement for LaO$_{0.5}$F$_{0.5}$BiS$_2$ (AP2). The obtained lattice parameters and reliability factor are $a$ = 4.0527 Å, $c$ = 13.3237 Å, $R_{wp}$ = 13.12%. The obtained atomic coordinates are summarized in Table 1. Figure 2(c-e) display the nominal $x$ dependence of the calculated lattice parameters of $a$, $c$ and $V$ for $x$ = 0 ~ 0.7 (AP1) and $x$ = 0.5 (AP2). The $a$ lattice parameter does not exhibit a remarkable F concentration dependence. The $c$ lattice parameter and volume ($V$) obviously decrease with increasing F concentration up to $x$ = 0.5. For $x$ >0.5, the changes in lattice constants are saturated, which indicates that the O/F solubility limit is near $x$ = 0.5.

*3.2 Magnetic susceptibility measurements for $LaO_{0.5}F_{0.5}BiS_2$.*

In this article, we show the physical property data for $x = 0.5$ because a superconducting transition was not observed for $x \leq 0.4$ and the best superconducting properties were obtained for $x = 0.5$. Figure 3(a, b) show the temperature dependence of magnetic susceptibility ($\chi$) from 15 to 2 K and an enlargement around the onset of the superconducting transition for $x = 0.5$ (AP1). The onset of $T_c$ is estimated to be ~2.7 K, and a large diamagnetic signal appears below 2.5 K. The calculated shielding volume fraction at 2 K is 11.0 %. Figure 3(c, d) show the temperature dependence of magnetic susceptibility from 15 to 2 K and an enlargement around the onset transition for $x = 0.5$ (AP2). The onset of $T_c$ is roughly estimated to be ~3 K, and a large diamagnetic signal appears below 2.5 K. The calculated shielding volume fraction at 2 K is 13.4 %. Although superconducting signals are observed in both Fig. 2(a) and 2(c), the $T_c$ and the shielding volume fraction were low. However, interestingly, both samples showed very slight drops at 10 K in the temperature dependence of magnetic susceptibility. Then, we expected that the maximum $T_c$ of this system is as high as 10 K. This implies that the electron doping level is not optimal, namely under-doped, due to the O/F solubility limit near $x = 0.5$. Therefore, we performed high pressure annealing to further increase the F concentration, because the high pressure synthesis technique is generally advantageous for obtaining samples with a smaller lattice. Figure 3(e, f) show the temperature dependence of magnetic susceptibility for $x = 0.5$ (HP). As we expected, very large shielding volume fraction as large as 100 % is observed. Furthermore, a dramatic enhancement of $T_c$ is observed. The onset of superconducting transition is above 10 K, and a large diamagnetic signal is observed below 8 K as shown in Fig. 3(f).

*3.3 Transport properties of $LaO_{0.5}F_{0.5}BiS_2$.*

Figure 4(a) shows the temperature dependence of resistivity for $x = 0.5$ (HP). Resistivity increases with decreasing temperature and a superconducting transition is observed at low temperatures. This behavior resembles the case of the carrier-doped band insulators, such as B-doped diamond[18,19] and the HfNCl family[20-22]. As shown in Fig. 4(b), the onset of the superconducting transition is observed at 10.6 K where the onset temperature ($T_c^{onset}$) is defined to be a temperature at which resistivity begins to decrease as indicated by the arrow. Figure 4(c) displays the temperature dependence of resistivity below 15 K under magnetic fields up to 5 T. To obtain a magnetic field – temperature phase diagram, we plotted the $T_c^{onset}$ and the zero-resistivity temperature ($T_c^{zero}$) with the respective applied fields in Fig. 4(d). The upper critical field is estimated to be 10 T using the WHH theory, which gives $\mu_0 H_{c2}(0) = -0.69 T_c (d\mu_0 H_{c2} / dT)|_{Tc}$.[23]

4. **Discussion**

Here we discuss the requirements of the evolution of bulk superconductivity in $LaO_{1-x}F_xBiS_2$. The F-concentration dependence of structural properties indicates that the O/F solubility limit seems to be near $x = 0.5$. A superconducting transition is observed for $x$ (nominal) $\geq 0.5$. However the shielding volume fraction in the magnetic susceptibility measurement is only 13.4 % for the best sample of $x = 0.5$ (AP2), which implies that the electron doping level is in the under-doped region of this system. Therefore, we performed the high pressure annealing to achieve further doping of F. Figure 5 shows the XRD profiles near the (102) and (004) peaks. The (004) peak slightly shifts to higher angles after high pressure annealing, indicating that the $c$ axis decreases. This change indicates that the F concentration increased by the high pressure annealing. As described in *3-1*, the $x = 0.5$ (AP2) sample contains small amount of $BiF_3$ and $LaF_3$. The F ions would be supplied from the impurity fluorides. With the fact that bulk superconductivity can be achieved by a slight increase of F concentration from $x = 0.5$, we guess the optimal doping level is slightly above $x = 0.5$. The estimated optimal doping level is comparable to that of $Bi_4O_4S_3$. Therefore, we will be able to achieve bulk superconductivity in $LaO_{1-x}F_xBiS_2$ by optimizing the carrier doping techniques or modifying the optimal doping level (band structure) by changing the spacer layer structures.

In conclusion, we have discovered a new type of $BiS_2$-based superconductor $LaO_{1-x}F_xBiS_2$ with a $T_c$ as high as 10.6 K. This superconductor has a layered structure analogous to the $BiS_2$-based $Bi_4O_4S_3$ superconductor, suggesting that the $BiS_2$ layer is a basic structure which provides new layered superconducting family like $CuO_2$ plane of cuprates and $Fe_2An_2$ layers of Fe-based family. Furthermore, we note that the spacer layer of $LaOBiS_2$ is almost the same as that of LaOFeAs, which is the first FeAs-based system. Thus, we will be able to discover many $BiS_2$-based superconductors on the basis of the analogy to FeAs-based compounds. We expect that the novel superconductors with the $BiS_2$ superconducting layers will open a new field in physics and chemistry of low-dimensional superconductors.

**Acknowledgment**

The authors would like to thank Dr. S. J. Denholme of National Institute for Materials Science (NIMS), Dr. H. Okazaki of NIMS, Dr. T. Yamaguchi of NIMS and Dr. H. Takatsu of Tokyo Metropolitan University for their experimental helps and fruitful discussion. This work was partly supported by Grant-in-Aid for Scientific Research (KAKENHI) and JST-EU-JAPAN project on superconductivity.

**Table. 1. Atomic coordinates for $LaO_{0.5}F_{0.5}BiS_2$ (HP).**

| site | $x$ | $y$ | $z$ | Occupancy |
|------|-----|-----|-----|-----------|
| La1  | 0.5 | 0   | 0.1015 | 1 |
| Bi1  | 0.5 | 0   | 0.6231 | 1 |
| S1   | 0.5 | 0   | 0.3657 | 1 |
| S2   | 0.5 | 0   | 0.8198 | 1 |
| O/F  | 0   | 0   | 0      | 0.5/0.5(Fixed) |

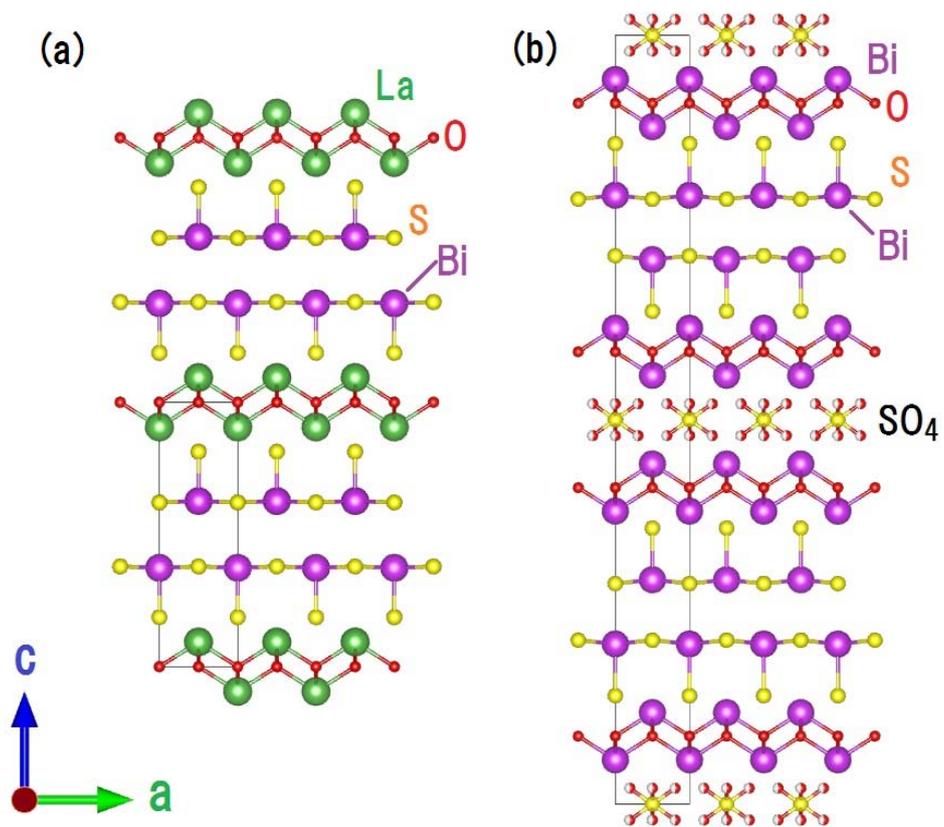

Fig. 1. (Color online) Crystal structure of (a)LaOBiS$_2$ and (b)Bi$_4$O$_4$S$_3$. The solid line indicates the unit cell.

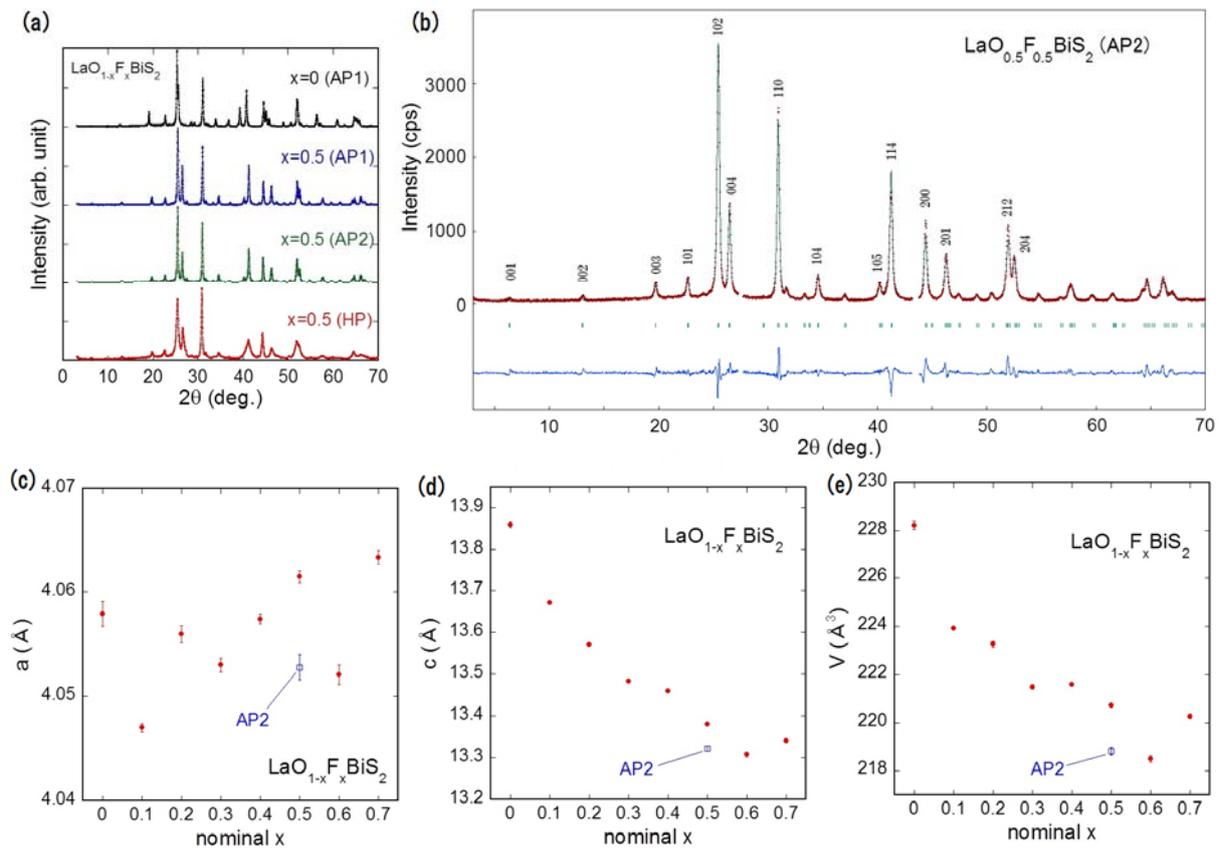

Fig.2. (Color online) XRD patterns and Crystal structural analysis of $LaO_{1-x}F_xBiS_2$. (a) XRD pattern for $x = 0$ and 0.5 synthesized by three different methods of AP1, AP2 and HP.   (b) XRD pattern and the result of Rietveld refinement for $LaO_{0.5}F_{0.5}BiS_2$ (AP2). The numbers are the Miller indices.   (c-e) Lattice parameters of $a$, $c$ and $V$ for $LaO_{1-x}F_xBiS_2$ (AP1) (red circles) and $LaO_{0.5}F_{0.5}BiS_2$ (AP2) (blue squares).

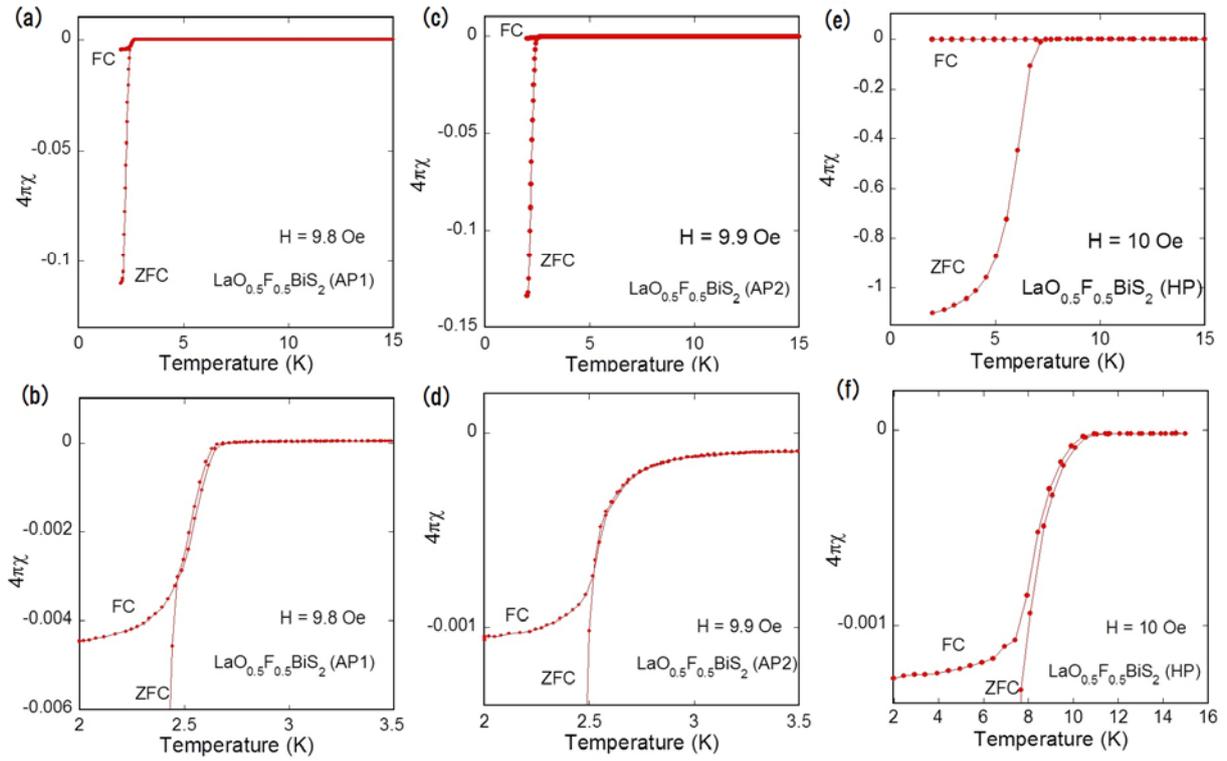

Fig. 3.   (Color online) Temperature dependence of magnetic susceptibility for LaO$_{0.5}$F$_{0.5}$BiS$_2$ synthesized by three methods of AP1, AP2 and HP.   (a, b) Temperature dependence of magnetic susceptibility for LaO$_{0.5}$F$_{0.5}$BiS$_2$ (AP1). (c, d) Temperature dependence of magnetic susceptibility for LaO$_{0.5}$F$_{0.5}$BiS$_2$ (AP2).   (e, f) Temperature dependence of magnetic susceptibility for LaO$_{0.5}$F$_{0.5}$BiS$_2$ (HP).

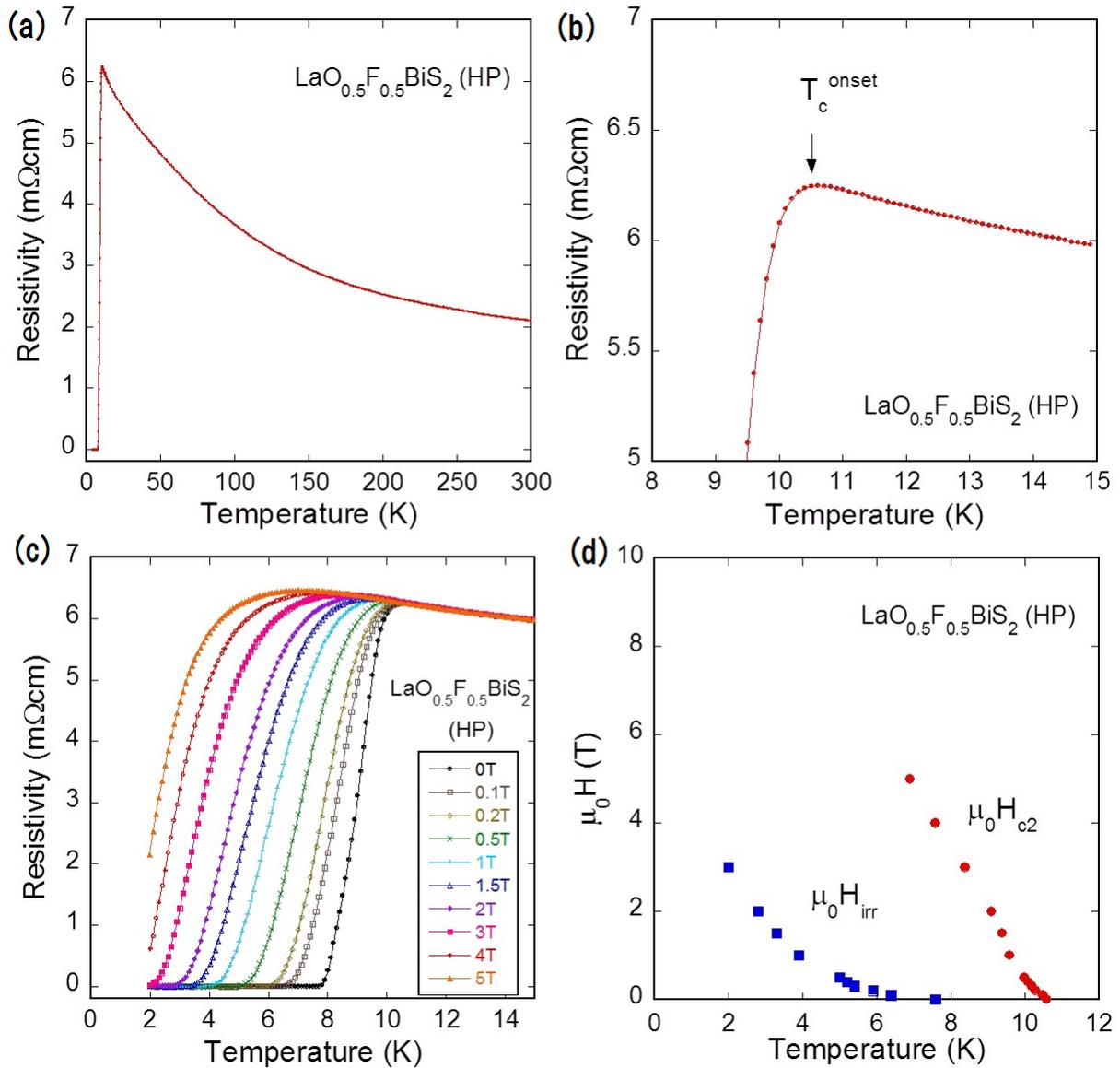

Fig. 4. (Color online) Transport properties of LaO$_{0.5}$F$_{0.5}$BiS$_2$ (HP). (a) Temperature dependence of resistivity for LaO$_{0.5}$F$_{0.5}$BiS$_2$ (HP). (b) Enlargement of superconducting transition for LaO$_{0.5}$F$_{0.5}$BiS$_2$ (HP) at 0 T. The onset of $T_c$ was estimated as indicated by an arrow. (c) Temperature dependence of resistivity for LaO$_{0.5}$F$_{0.5}$BiS$_2$ (HP) under magnetic fields up to 5 T. (d) Magnetic field – Temperature phase diagram of LaO$_{0.5}$F$_{0.5}$BiS$_2$ (HP).

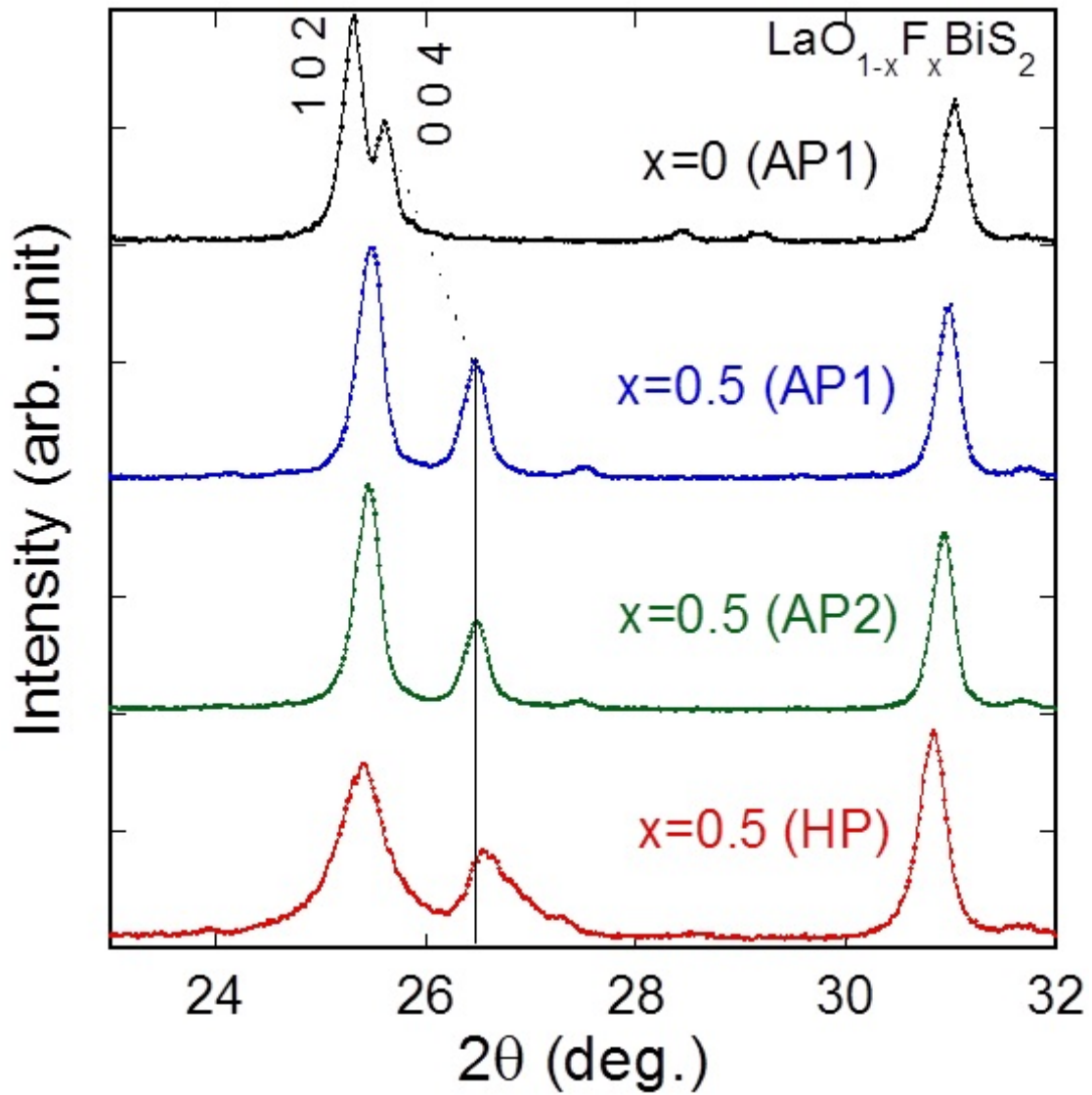

Fig. 5. (Color online) Comparison of (102) and (004) peaks between the sample preparation methods. The dashed and solid lines are the guidelines to clarify the shift of the (004) peaks.